\documentclass[11pt]{article}
\usepackage{times}
\usepackage{geometry}
\usepackage[utf8]{inputenc}
\usepackage{enumitem,amssymb}
\usepackage{ragged2e}
\newlist{thematic}{itemize}{8}
\setlist[thematic]{label=$\square$}
\usepackage{pifont}
\usepackage{enumitem}
\setlist[itemize]{leftmargin=*}
\usepackage{fancyhdr}
\usepackage[USenglish]{babel}
\usepackage[utf8]{inputenc}
\usepackage[T1]{fontenc}
\usepackage{epsfig}
\usepackage{wrapfig}
\usepackage{color}
\usepackage[vflt]{floatflt}
\usepackage{longtable}
\usepackage{caption}
\usepackage{jheppub}   
\usepackage[dvipsnames,svgnames,x11names]{xcolor}
\usepackage[all]{hypcap} 
\usepackage{amssymb,amsmath}
\usepackage{longtable}
\usepackage{amssymb}
\usepackage{amsmath, xspace}
\usepackage{graphics,graphicx} 
\usepackage{tikz}
\usetikzlibrary{arrows.meta} 
\usepackage{rotating}
\usepackage{color}
\usepackage{xcolor}
\usepackage{multirow}
\usepackage{subfigure}
\usepackage[outercaption]{sidecap}
\sidecaptionvpos{figure}{c}

\usepackage{enumitem}
\setlist[enumerate]{itemsep=0pt, parsep=0pt}
\usepackage{titlesec}
\titlespacing*{\section}{0pt}{3pt}{4pt} 
\titlespacing*{\subsection}{0pt}{6pt}{2pt}

\let\oldbibliography\thebibliography
\renewcommand{\thebibliography}[1]{%
  \oldbibliography{#1}%
  \setlength{\itemsep}{0pt}
  \setlength{\parskip}{0pt}
}

\definecolor{DarkGreen}{rgb}{0.0, 0.3, 0.0}
\definecolor{purple}{rgb}{0.5, 0.0, 0.5}
\definecolor{red}{rgb}{1, 0.0, 0.0}
\definecolor{green}{rgb}{0, 1.0, 0.0}

\def\3he{$^3{\rm He}$}

\def\lsim{\mathrel{\lower2.5pt\vbox{\lineskip=0pt\baselineskip=0pt
           \hbox{$<$}\hbox{$\sim$}}}}

\def\gsim{\mathrel{\lower2.5pt\vbox{\lineskip=0pt\baselineskip=0pt
           \hbox{$>$}\hbox{$\sim$}}}}

\title{\vspace{2cm}\textbf{
The Dynamics of the Milky Way: Unveiling the 6D Skeleton of Star Formation in the 2040s} \\
\vspace{1cm} 
ESO call for White Paper 2025}
\author{Loredana Prisinzano$^{1}$, 
Germano G. Sacco$^{2}$, 
Francesco Damiani$^{1}$,
Amelia Bayo$^{3}$,
Salvatore Sciortino$^{1}$, \\
Marco Tarantino$^{4}$, 
Rosaria Bonito$^{1}$,
Fatemeh Zahra Majid$^{5}$}
\date{\vspace{0.2cm}$^{1}$INAF-Oss. Astronomico di Palermo,
$^{2}$INAF-Oss. Astronomico di Arcetri,
$^{3}$ESO-Garching, Germany, 
$^{4}$Palermo University, 
$^{5}$INAF-Oss. Astronomico di Capodimonte\
}
\begin{document}

\maketitle
\thispagestyle{empty}
\begin{figure}[!b]
    \centering
    \begin{tikzpicture}
        \node[anchor=south west, inner sep=0] (image) at (0,0) {
            \includegraphics[width=0.80\textwidth,clip,trim=0 0 0 42]{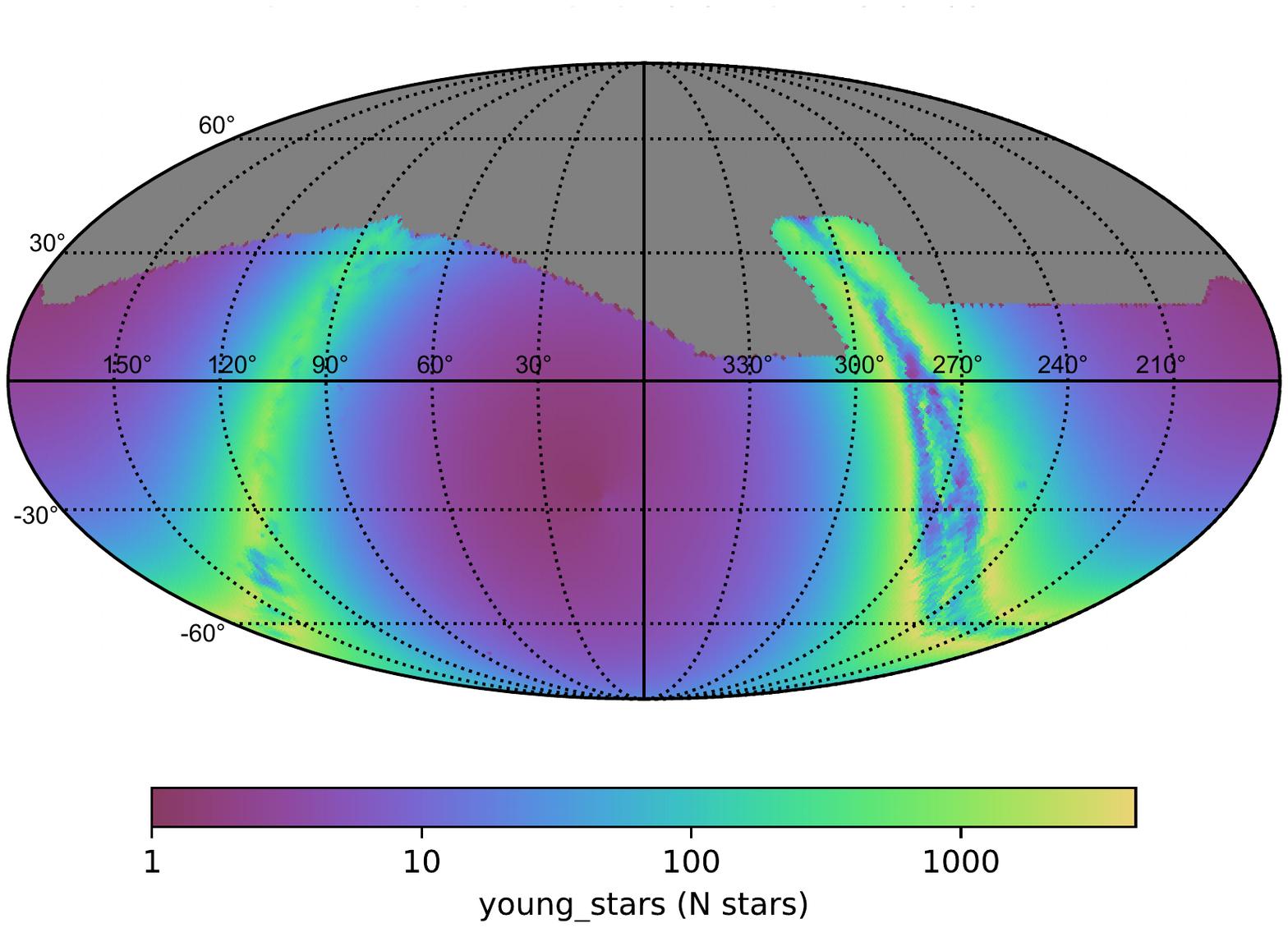}
        };

        \begin{scope}[x={(image.south east)}, y={(image.north west)}]
            

            \draw[->, white, line width=1.5pt, -{Stealth[length=3mm]}] (0.82, 0.75) -- (0.75, 0.60);
             \draw[->, white, line width=1.5pt, -{Stealth[length=3mm]}] (0.35, 0.75) -- (0.28, 0.60);           
            \node[white, font=\bfseries\sffamily\small, anchor=west, align=left] at (0.72, 0.77) {Galactic\\Plane};
            \node[white, font=\bfseries\sffamily\small, anchor=west, align=left] at (0.35, 0.77) {Galactic\\Plane};
            
        \end{scope}
    \end{tikzpicture}

    \captionsetup{labelformat=empty}
    \caption{\small Fig.1 Map of the predicted number of stars younger than 10\,Myr down to 0.3\,$M_\odot$ per HEALPix ($N_{side} = 64$) shown in equatorial Mollweide projection. The counts are computed following the metric used in \citet{pris23}, incorporating the 3D dust map and crowding effects based on the Rubin LSST Opsim. The map represents the sources detectable down to the limiting magnitudes corresponding to an accuracy level of $\sigma = 5$ in the $g, r, i$ bands over the full 10-year survey. Adapted from Prisinzano et al. \citep{pris23}.}
\end{figure}

\newpage


\section*{Key Science Questions}
The Milky Way (MW) is our unique laboratory to test star formation theories at the level of individual stars, serving as the Rosetta Stone to interpret extragalactic observations. The proposed White Paper focuses on the following key questions regarding the structure and evolution of the MW  traced by young stellar populations:
\textbf{Q1.} 
How do large-scale dynamical instabilities, like warps and vertical waves, drive the  star formation of the Galactic thin disk?
\textbf{Q2.} 
Do star-forming regions form stochastically, driven by local self-propagating feedback, or are they triggered by a common dynamical process acting on Galactic scales? Do internal feedback loops and external interactions tend to sustain or quench star formation in the MW?
\textbf{Q3.} Are the clustered star-forming regions the only environments where stars form, or can stars also form in more diffuse structures such as the stellar strings?
\section{State of the Art}
Our spiral Galaxy can serve as a benchmark for studying other spiral galaxies, since we can observe the full and detailed distribution of its star, gas and dust components. The spatial distribution of young stars in the MW is particularly important for defining its morphology, as young stellar objects (YSOs) are known to be located in the main skeleton shaped by the spiral arms.
YSOs are excellent tracers of the MW’s structure, as they directly reflect the physical processes involving gas and dust in the Galactic disc. However, the detailed model of the Galactic architecture is still poorly constrained, as, our internal vantage point makes studying the global structure inherently more challenging than observing external spiral galaxies from the outside. 

With the advent of Gaia, large scale structures such as, for example, the Radcliffe wave \citep{alve20,koni24},
the Local Arm/Split \citep{lall19},
the Sagittarius Arm \citep{kuhn21} and
the Cepheus Spur \citep{pant21}, have been found.
In addition, it has been revealed that the Galactic disc is a complex dynamical system characterized by large-scale non-axisymmetric features, including a central bar, spiral arms, and a prominent outer warp \citep{scho18}. The disc has been revealed to be out-of-equilibrium, exhibiting vertical North-South asymmetries and kinematic substructures such as the phase spiral, likely induced by perturbations from the Sagittarius dwarf galaxy \citep[e.g.][]{blan21}. While the kinematic signature of the warp has been successfully modeled as a precessing structure, simulations suggest that satellite interactions can also excite coherent vertical corrugations and bending waves that propagate through the stellar disc \citep{pogg25}.


These recent findings demand a  step forwards from the era of static geometry to one of dynamic evolution. In fact, the challenge is not merely to locate the spiral arms but to understand how they interact with the Galaxy's global oscillations. We now have hints of a vertical wave propagating towards the outer disk, superimposed on a precessing warp. It remains unknown how these propagating waves, likely excited by satellite interactions or internal mechanisms, affect star formation and the shape of spiral arms in the MW.

\section{Scientific challenges}
Building a spatial (X,Y, Z) and kinematic (proper motions and radial velocities) 6D map of the MW’s young structures is crucial to unveil the present-day morphology of our Galaxy, including the definition of its spiral arms and large-scale 
substructures such as waves, spurs, and warps. 
While the Gaia mission has revolutionized our view of the solar neighborhood, only a few attempts—sometimes yielding conflicting results—have been made to define a model of the MW’s spiral structure, in particular for the Local, Perseus, and Sagittarius-Carina arms \citep{pogg21}.

The far side of the MW, including the Scutum-Crux and the Norma arms toward the Galactic Center, and the Outer arm, toward 
the Galactic anticenter,  remains largely uncertain. In fact, current knowledge of the spiral arm geometry relies  on sparse tracers such as masers, Cepheids,  OB stars and young giant stars.
However, these populations are limited in number and often affected by selection biases. 
In the next decade, the exceptional deep photometry from Vera C. Rubin Observatory Legacy Survey of Space and Time (LSST) and Nancy Grace Roman Telescope will allow
us to make significant advances in understanding the MW’s structure,
by including also young low-mass ($<2\,M_\odot$) populations down to M-type stars (Fig. 1). These latter
represent the dominant component of stellar populations (over 80\%; \citep{lada06}) and are therefore crucial, as well as statistically favorable, for tracing the morphology of the full populations formed from giant molecular clouds. The years after 2040 could mark the era of \textbf{\textit{GaiaNIR}}, a proposed ESA Large mission designed to survey $\sim$12 billion stars, using NIR astrometry to penetrate dust-obscured regions and finally reveal the Galaxy's hidden complexity \citep{hobb24}.

The main challenge in the next few decades is not merely to reconstruct the static 3D morphology of the MW's unexplored side, but to unveil its \textbf{dynamical nature}. 

 However, to fully determine how large-scale vertical instabilities — such as warps and propagating waves— dynamically couple with the disk, astrometry alone is insufficient. It is therefore crucial to establish a corresponding ground-based spectroscopic facility capable of complementing \textit{GaiaNIR} with precise radial velocities. This synergy is mandatory to complete the 6D phase-space map in the highly extincted inner regions, allowing us to trace not only the full dynamical evolution of the spiral structure
but also other key large-scale processes.

The second scientific challenge is to consider \textbf{stellar ages}. By reconstructing the age distribution of these stars across the unexplored disk, the Star Formation History (SFH) of large-scale structures can be constrained, assuming 
that spatially coherent, coeval structures are likely driven by a common dynamical origin.
Recent analyses, such as, for example, the oscillatory motion and radial drift of the Radcliffe Wave \citep{koni24}, suggest that the Galactic gravitational potential plays an important role in shaping the dynamics of large-scale star-forming structures.
 This observational evidence is supported by recent cosmological simulations showing that the Cluster Formation Efficiency (CFE) and the initial cluster mass function  depend on the gas pressure and surface density of the large-scale galactic environment \citep{grud23}. 
 This highlights a physical link in which evolving large-scale conditions must be taken into account when interpreting the properties of star formation on small scales.

 Characterizing entire coeval coherent  populations will be therefore crucial to understand if the star formation is mainly triggered by local photoionisation-driven feedback from OB stars or supernovae explosion,  or if  large-scale effects,
such as perturbation waves caused by the Galaxy’s rotation or gravitational potential,
previously unconsidered, might also play a dominant role in the star formation process.

By analyzing the age distribution, we can discriminate  different formation scenarios. For example episodic 'bursty' events could correspond to high-efficiency formation episodes driven by extreme environmental pressures (potentially merger-induced). On the other hand, recent hydrodynamical simulations of MW-like galaxies (e.g. \citep{barb25}) demonstrate that a continuous, self-regulated baryon cycle, driven by galactic fountains and radial inflows, can sustain a constant Star Formation Rate over gigayears without the need for major external mergers. This supports the scenario where continuous age distributions in the disk point towards secular evolution driven by smooth gas accretion. Disentangling these scenarios requires understanding the dual role of feedback on intermediate scales: does the energy injection from young stars primarily disperse the parental cloud (quenching), or does it compress surrounding material to trigger further star formation (sustaining)?

The third scientific challenge lies in determining whether star formation occurs exclusively in dense, gravitationally bound clusters, or whether it also takes place in more diffuse environments. Recent Gaia data have revealed the existence of extended (100–400 pc), low-density kinematic structures, commonly referred to as “stellar strings” (e.g., Theia groups), whose physical origin remains uncertain \citep[e.g.][]{koun19,zuck22}. What we need to understand is whether these strings represent (i) the dispersed remnants of originally dense clusters or (ii) genuinely primordial, coherent structures where star formation occurred in situ under low-density conditions \citep{wrig23,koun20}. Establishing whether strings trace an alternative mode of star formation would imply that this process spans a broader continuum of environmental densities \citep{krui12,mire24} than traditionally assumed, challenging the classical ‘clustered’ paradigm\citep[e.g.][]{lada03}. 

\section{Technology Developments and Data-Handling Requirements}
While the extraordinary depth of upcoming photometric facilities, such as Gaia DR5, Rubin LSST, the Roman Telescope, and, hopefully, GaiaNIR will revolutionize our perspective of the MW and allow us to unveil its elusive far side, no corresponding spectroscopic facilities are currently planned to match this capability in the 2040s.

A 12-m class seeing limited Wide-field Spectroscopic Telescope  (FoV$\sim$ 3-5\,deg$^2$) to perform spectroscopic surveys of low mass  (2-0.2\,M$_\odot$) young ($\lesssim 100$\,Myr) stars will crucially allow us to study stellar populations in the peak of the mass function.
This facility should be equipped  with 
a multi-object spectrograph (MOS) with thousands fibers, 
and resolution $\gtrsim4000$ operating in the band 500-1800\,nm, to measure both radial velocities with a precision of $\sim$1-2 km/s, and key stellar parameters, such as effective temperatures and surface gravities, essential ingredients to infer accurate masses and ages.
 However, the high extinction caused by interstellar dust makes optical observations alone very hard, especially in the Galactic center direction. 
  To penetrate the heavy extinction of the Galactic Plane and reach the Far Side, 
  the extension into the Near-Infrared domain—specifically covering the J and H bands—is
mandatory.


To quantify the gain provided by this instrument over facilities like 4MOST, we considered a 10\,Myr solar metallicity isocrone 
at increasing distances and extinctions. Since the 4MOST limit of young stars is $G_{lim} \approx 18.5$, spectroscopic coverage for the bulk of the stellar population (down to late M-type stars) is confined to the solar neighborhood ($d \lesssim 500$\,pc). Assuming a limiting magnitude $G_{lim} \approx 22$,   such  transformative instrument will enable the detection of low-mass tracers (down to $M \sim 0.5\,M_{\odot}$) out to $2$\,kpc, at extinction $A_G \lesssim 3$. 
In addition, we can detect down to  FG-type stars out to 3-4\,kpc, at 
$A_G \lesssim 5$.
  Synergy with GaiaNIR, Rubin LSST and Roman Telescope will enable efficient 
 identification of
 YSOs candidates \citep[e.g.][]{pris23,boni23}.


\bibliographystyle{aa} 
\bibliography{references} 
\end{document}